\documentclass[twoside]{article}
\usepackage{pmart,fleqn,epsfig}
\usepackage{amssymb,amsmath}

\begin{document}
%
%
\title{Unidirectional Charge Instability of \\  the $d$-wave RVB Superconductor}

\author{Marcin Raczkowski$^{a}$, Manuela Capello$^{b}$ and Didier Poilblanc$^{b}$}

%
%
\address{$^a$Marian Smoluchowski Institute of Physics, Jagellonian
             University,\\ Reymonta 4, PL-30059 Krak\'ow, Poland \\
         $^{b}$Laboratoire de Physique Th\'eorique UMR5152, CNRS,
         F-31062 Toulouse, France }

\maketitle                


\begin{abstract}
Starting from  a uniform $d$-wave superconducting phase we study the energy cost
due to imposed unidirectional defects with a vanishing pairing amplitude. 
Both renormalized mean-field theory and variational Monte Carlo calculations
within the $t$-$J$ model yield that the energies of inhomogeneous and uniform
phases are very close to each other.
This suggests that small perturbations in the microscopic Hamiltonian,
might lead to inhomogeneous superconducting phases in real materials as
observed in recent scanning tunneling microscopy on Ca$_{2-x}$Na$_x$CuO$_2$Cl$_2$. 

\end{abstract}

\section{Introduction} 

Recent progress in spectroscopic techniques has provided a wide variety of
interesting data concerning electronic states of the high-$T_c$
superconductors. For example, scanning tunneling microscopy (STM) on 
different cuprate families Ca$_{2-x}$Na$_x$CuO$_2$Cl$_2$ and
Bi$_2$Sr$_2$Dy$_{0.2}$Ca$_{0.8}$Cu$_2$O$_{8+\delta}$, has revealed
\emph{short-range} unidirectional charge domains coexisting with inhomogeneous
$d$-wave superconductivity \cite{Koh07}. In particular, it has been found,
that the doped holes primarily enter oxygen sites leading to a
\emph{bond-centered} charge pattern with a period of four lattice spacings. 
Motivated by this result, we have recently shown \cite{Rac07}, that such a
charge order might be naturally interpreted in terms of a valence bond crystal 
\cite{Sach08}, i.e., paramagnetic phase with both spatially varying bond
charge hopping and short-range antiferromagnetic (AF) correlations. 
In this case, an inhomogeneous antiphase domain resonating valence bond 
($\pi$DRVB) phase was obtained by assuming a $\pi$-phase shift in the 
superconducting (SC) order parameter across domain walls (DWs).  
While the antiphase solution is particularly intriguing since, in contrast to 
its inphase counterpart, offers a simple explanation of the suppression 
of the effective interlayer Josephson coupling observed in some stripe-ordered
high-$T_c$ compounds \cite{Oga02,Ber07}, both types of the modulation of the 
SC order parameter are the subject of intense ongoing studies \cite{Voj1,Bar08,Manu,Voj2}.
Therefore, in this paper we shall study  the energy cost due to imposed
defects with a vanishing pairing amplitude (no a $\pi$-shift is assumed across
the DWs) and compare the resulting charge modulation 
with the corresponding one found in the $\pi$DRVB phase.

\section{Model and the approach}

We investigate a $t$-$J$ model Hamiltonian,   
\begin{equation}
{\cal H}= - t \sum_{\langle ij\rangle,\sigma}
     ({\tilde c}^{\dag}_{i\sigma}{\tilde c}^{}_{j\sigma} + h.c.)
      + J\sum_{\langle ij\rangle} {\bf S}_i \cdot {\bf S}_j,
\label{eq:tJ}
\end{equation}
where ${\tilde c}^{\dag}_{i\sigma}=(1-n_{i,-\sigma})c^{\dag}_{i\sigma}$ 
is the Gutzwiller projected electron operator and use a renormalized 
mean field theory (RMFT) in which the local constraints of no doubly 
occupied sites are replaced by statistical Gutzwiller weights 
$g_{ij}^t$ ($g_{ij}^J$) for hopping (superexchange) processes, 
respectively \cite{RVB}. Hence the mean-field Hamiltonian reads,
\begin{align}
\label{eq:H_MF}
              H_{MF} &= - t\sum_{\langle ij\rangle,\sigma} g_{ij}^t
                  (c^{\dagger}_{i,\sigma}c^{}_{j,\sigma}+h.c.)
                   -\mu\sum_{i,\sigma}n_{i,\sigma}\nonumber \\
                  &-\frac{3}{4} J \sum_{\langle ij\rangle,\sigma}g_{ij}^J
                  [(\chi_{ji}c^{\dagger}_{i,\sigma}c^{}_{j,\sigma}
                  + \Delta_{ji}c^{\dagger}_{i,\sigma}c^\dagger_{j,-\sigma}
                  + h.c.) -|\chi_{ij}|^2  -|\Delta_{ij}|^2],
\end{align}
with the Bogoliubov-de Gennes self-consistency conditions for the bond-
$\chi_{ji}=\langle c^\dagger_{j,\sigma}c^{}_{i,\sigma}\rangle$ and pair-order
$\Delta_{ji}=\langle c^{}_{j,-\sigma}c^{}_{i,\sigma}\rangle
            =\langle c^{}_{i,-\sigma}c^{}_{j,\sigma}\rangle$
parameters in the unprojected state. We consider here the 
so-called modified Gutzwiller factors, 
\begin{align}
  g_{ij}^J &=\frac{4(1-n_{hi})(1-n_{hj})}
                   {\alpha_{ij}+8n_{hi}n_{hj}\beta_{ij}^{-}(2)
                   +16\beta_{ij}^{+}(4)},
\label{eq:GJ} \\
  g_{ij}^t &=\sqrt{\frac{4n_{hi}n_{hj}(1-n_{hi})(1-n_{hj})}
                   {\alpha_{ij}+8(1-n_{hi}n_{hj})|\chi_{ij}|^2
                   +16|\chi_{ij}|^4}},
\label{eq:Gt}
\end{align}
where $\alpha_{ij}=(1-n_{hi}^2)(1-n_{hj}^2)$,
$\beta_{ij}^{\pm}(n)= |\Delta_{ij}|^n\pm|\chi_{ij}|^n$ while $n_{hi}$ are 
local hole densities. By including the effects of the 
nearest-neighbor correlations $\chi_{ij}$ and $\Delta_{ij}$  they are known to
give a better agreement with a more accurate Variational Monte Carlo (VMC) 
technique \cite{Rac07}. 
Hereafter, we shall assume a typical value $t/J=3$ and fix the doping level
$x=1/8$. Finally, using unit cell translation symmetry \cite{Rac06}, RMFT
calculations were carried out on large $256\times 256$ clusters at a low 
temperature $\beta J=500$ approaching thermodynamic limit.

\section{Results and discussion}

\begin{figure}[t!]
\begin{center}
\resizebox{0.85\linewidth}{!}{
\includegraphics*{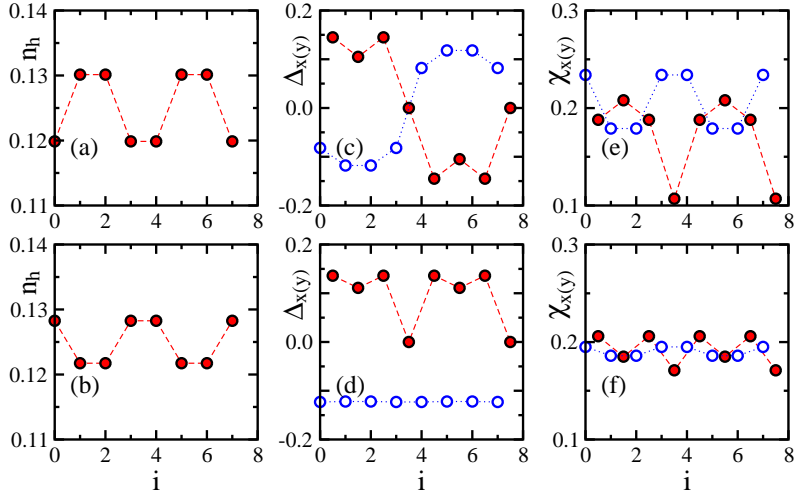}}
\end{center}
  \caption{
  (a,b) Hole density $n_{hi}$ and variational parameters: 
  (c,d) $\Delta_{i,i+\alpha}$ as well as 
  (e,f) $\chi_{i,i+\alpha}$  found in the $\pi$DRVB (top) and DRVB (bottom) phase.
  Solid (open) circles in panels (c-f) correspond to the $x$ ($y$) direction, 
  respectively.
}
  \label{fig1}
\end{figure}

In Fig.~\ref{fig1} we show the hole profiles as well as the values of the 
bond- and pair-order parameters across the unit cell found in the $\pi$DRVB
(top) and inphase DRVB (bottom) state. The obtained modulations clearly
reflect the competition between the superexchange energy $E_J$ and kinetic 
energy $E_t$ of doped holes. However, a detailed charge profile depends on 
the assumed type of the SC order parameter. On the one hand, suppression of
the pair-order amplitude $\Delta_{ij}$ along the DWs automatically involves 
a deviation of the bond-order parameter $\chi_{ij}$ from the value found in the areas 
with finite $\Delta_{ij}$. Remarkably, the deviation is particularly strong  
in the case of the antiphase SC order parameter. On the other hand, 
the absence of the $\pi$ shift across the stripe boundary in the DRVB phase 
allows the system (as confirmed by the VMC  method \cite{Manu}) to avoid 
a reduction of $\Delta_{ij}$ on the adjacent vertical bonds which 
remains almost intact. Therefore, the charge redistributes from the hole rich areas 
with enhanced $\Delta_{ij}$ in the $\pi$DRVB phase \cite{Rac07}, towards DWs 
with vanishing $\Delta_{ij}$ in the DRVB state (see Fig.~\ref{fig1}).

\begin{figure}[t!]
\begin{center}
\resizebox{0.85\linewidth}{!}{
\includegraphics*{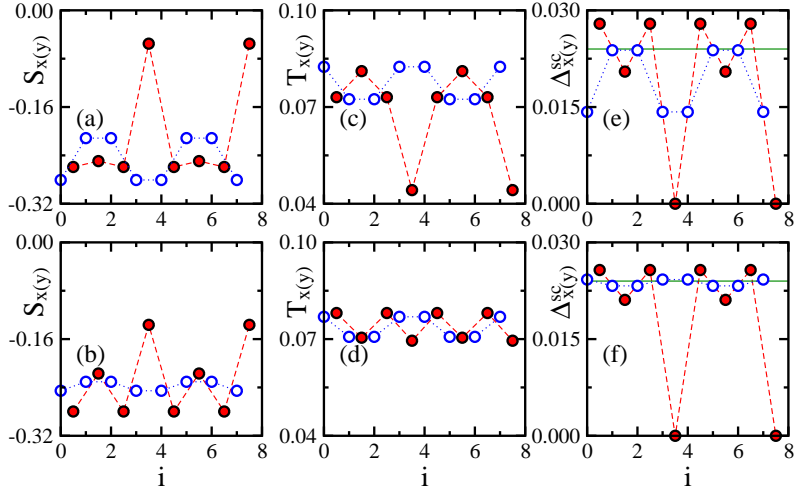}}
\end{center}
  \caption{
  (a,b) Spin correlation ${\mathcal S}_{i}^{\alpha}$, 
  (c,d) bond charge ${\mathcal T}_{i}^{\alpha}$, and 
  (e,f) SC order parameter $\Delta_{i\alpha}^{\rm SC}$ found in 
  the $\pi$DRVB (top) and DRVB (bottom) phase.
  Solid (open) circles correspond to the $x$ ($y$) direction, respectively; 
  solid line in panels (e,f) depicts the SC order parameter in the uniform 
  $d$-wave RVB phase. 
}
  \label{fig2}
\end{figure}

In order to appreciate better the reason of a different charge profile in both
phases we show in Fig.~\ref{fig2}(a-d) the corresponding short-range AF correlations,
\begin{equation}
{\mathcal S}_i^{\alpha}=
-\frac{3}{2}g_{i,i+\alpha}^{J}(|\chi_{i,i+\alpha}|^2+|\Delta_{i,i+\alpha}|^2),
\end{equation}
with $\alpha=\{x,y\}$, as well as bond charge hopping,
\begin{equation}
{\mathcal T}_i^{\alpha}=2g_{i,i+\alpha}^{t}Re\{\chi_{i,i+\alpha}\}, 
\end{equation}
across the unit cell. 
Here one finds that a local reduction of the SC order
parameter (and the concomitant strong suppression of the superexchange energy
on the related bonds) enables, in the $\pi$DRVB phase, a large bond charge 
hopping along the DWs as in the usual stripe scenario \cite{Rac07}.  
In fact, it also determines the actual hole profile arranged in the way which 
minimizes the loss of the superexchange energy at the DWs. 
This can be easily accomplished by expelling the holes and strengthening
locally the corresponding $g_{ij}^J$ factors.   
In contrast, small modulation of $\chi_{ij}$ in the DRVB phase results 
in a much weaker, with respect to the $\pi$DRVB one, modulation of 
both the spin correlations and bond-charge hopping. Consequently, 
the system does not have to further improve the superexchange energy 
at the defect lines but it rather tries to regain some kinetic energy 
released on the broken RVB bonds. This is reached by adjusting the hole 
profile and attracting the holes to the DWs which enlarges locally 
renormalization factors $g_{ij}^{t}$ . 
As a result, the DRVB phase has a very good kinetic energy being even 
slightly better than that of the uniform $d$-wave RVB phase (see Table~I). 
\begin{table}[b!]    
  {\small 
    \rightline{TABLE I} 
    \noindent
    RMFT kinetic energy $E_t$, magnetic energy $E_J$, and free energy $F$
    as well as VMC energy $E_{\rm VMC}$ of the locally stable phases: 
    $\pi$DRVB, DRVB, and $d$-wave RVB one at $x=1/8$. 
\smallskip
 \begin{center}
    \begin{tabular}{c|c|c|c|c}
      \hline                
      \hline
     phase &  $E_{t}/J$   &  $E_{J}/J$   & $F/J$ &  $E_{\rm VMC}/J$   \\
      \hline
  $\pi$DRVB&  $-$0.8719   &  $-$0.4518  &  $-$1.3237 & $-$1.3359\\
       DRVB&  $-$0.8871   &  $-$0.4662  &  $-$1.3533 & $-$1.3647\\
        RVB&  $-$0.8863   &  $-$0.4784  &  $-$1.3647 & $-$1.3669
    \end{tabular}
  \label{table}
 \end{center}
}
\end{table}
Let us point out, however, that even though both the RMFT and VMC methods 
predict exactly the same hole profiles in the DRVB phase (as well as its
remarkably good energy), a discrepancy appears concerning kinetic 
energy gain at the DWs, strongly enhanced in the VMC method 
\cite{Manu}. The difference simply follows from the fact that in the RMFT 
both the short-range AF correlations and bond-charge hopping are
$\propto\chi_{ij}$. Hence its suppression involves a reduction of both the 
energy contributions unless the system is disposed towards a strong phase 
separation so that they can be further modified by the Gutzwiller 
factors \cite{SFP}.

Finally in order to discuss the SC properties of our inhomogeneous phases 
we plot in Fig.~\ref{fig2}(e,f) the modulus of SC order parameter, 
\begin{equation} 
\Delta^{\rm SC}_{i\alpha}=g_{i,i+\alpha}^{t}|\Delta_{i,i+\alpha}|,
\end{equation} 
across the unit cell. One of the key qualitative differences between 
the $\pi$DRVB and its inphase counterpart is evident in this
figure. Namely, while the SC order parameter deviates, in the regions between
defect lines, only slightly in both states from the value found in 
the uniform $d$-wave RVB phase, the absence of the $\pi$ shift across the
stripe boundary in the DRVB phase decouples the horizontal and 
vertical bonds constituting DWs. Therefore, in contrast to the
$\pi$DRVB phase, the latter retain the value of the SC order parameter of 
the uniform state.

\section{Summary and conclusions} 

In this paper we have studied two possible modulations of the SC order 
parameter across the DWs: inphase and antiphase. Remarkably, 
we have found that the energy of the unidirectional modulated phases 
(especially of the inphase configuration) approaches the energy of 
the uniform $d$-wave RVB superconductor. In fact, the energy difference might 
be further reduced by the tetragonal lattice distortion that often appears 
in the high-$T_c$ compounds \cite{Manu}.  
We conclude therefore that the $d$-wave RVB phase is capable of efficient 
minimizing the energy cost due to unidirectional defects with 
broken RVB bonds which in turn might induce the charge 
modulation similar to that observed in the STM experiments \cite{Koh07}.

\section*{Acknowledgments}
M.R. acknowledges support from the Foundation for Polish Science (FNP) and 
from Polish Ministry of Science and Education under Project 
No. N202 068 32/1481. M.C. and D.P. acknowledge the Agence Nationale
de la Recherche (France) for support.

\end{document}